\documentclass{llncs}
\usepackage{amsmath}
\usepackage{graphicx}

\title{Proof-Carrying Hardware via IC3} 
\author{Tobias Isenberg \and Heike Wehrheim}
\institute{Universit\"at Paderborn, Institut f\"ur Informatik, \\ 33098
  Paderborn, Germany\\
{\small $\{$isenberg,wehrheim$\}$@mail.upb.de}}

\begin{document}

\maketitle

\begin{abstract}  Proof-carrying hardware (PCH) is an approach to achieving safety of dynamically reconfigurable hardware, transferring the  idea of proof-carrying code to the hardware domain. Current PCH approaches are, however, either limited to combinational and  bounded unfoldings of sequential circuits, or only provide semi-automatic proof generation.  We propose a new approach to PCH which employs IC3 as proof generator, making automatic PCH applicable to sequential circuits in their full generality. We demonstrate feasibility of our approach by showing that {\em proof validation} is several orders of magnitude faster than original proof generation while (most often) generating smaller proofs than current PCHs.
\end{abstract} 

\section{Introduction}

Proof-carrying hardware (PCH)  \cite{DrzevitzkyBoundedSequEquiv,LoveIP} was introduced to ensure safety of dynamically reconfigurable hardware in situations where the hardware provider is not fully trusted and a complete verification is infeasible due to time restrictions.
To this end, PCH is applying  the principle of proof-carrying code to hardware modules. 
Proof-carrying code (PCC) as introduced by Necula \cite{Necula97} aims at the safe execution of code written by untrusted code producers and shipped via untrusted mediums to code consumers (e.g., mobile devices).
Its goal is to convince the consumers of the safety of the code. 
To this end, code producers attach safety proofs to their code, and consumers validate these proofs.
Such techniques should be tamper-proof: malicious modifications of the code or the proof are to be detected. 
Furthermore, proof validation for the consumer should be significantly faster than proof generation by the producer ({\em speed-up}). 
PCH aims at achieving these effects for reconfigurable hardware.

Current approaches to PCH are, however, limited to combinational circuits or bounded unfoldings of sequential circuits  \cite{DP11,DrzevitzkyBoundedSequEquiv}, or use an interactive prover (Coq) for proof generation \cite{LoveIP} and thus cannot provide an automatic PCH technique.

In this paper, we propose an approach for PCH based on the IC3 \cite{Bradley11} algorithm. 
For a given arbitrary safety property and circuit, IC3 computes an inductive strengthening of the property showing its validity (or generates a counterexample). 
IC3 is fully automatic and most suitable for sequential circuits, ruling out the need to unroll the transition relation.
At the consumer, proof validation proceeds by checking whether the proof is indeed an inductive strengthening of the safety property for the shipped hardware reconfiguration.

We show that this approach is tamper-proof
and experimentally evaluate it on sequential multipliers
and  some of the benchmarks of the
hardware model checking competition 2013.
The experiments show a significant {\em speed-up}, thus making our approach an ideal candidate for PCH. 
\section{Concept}

\begin{figure}[htp]
\centering
	\includegraphics[width=0.9\textwidth]{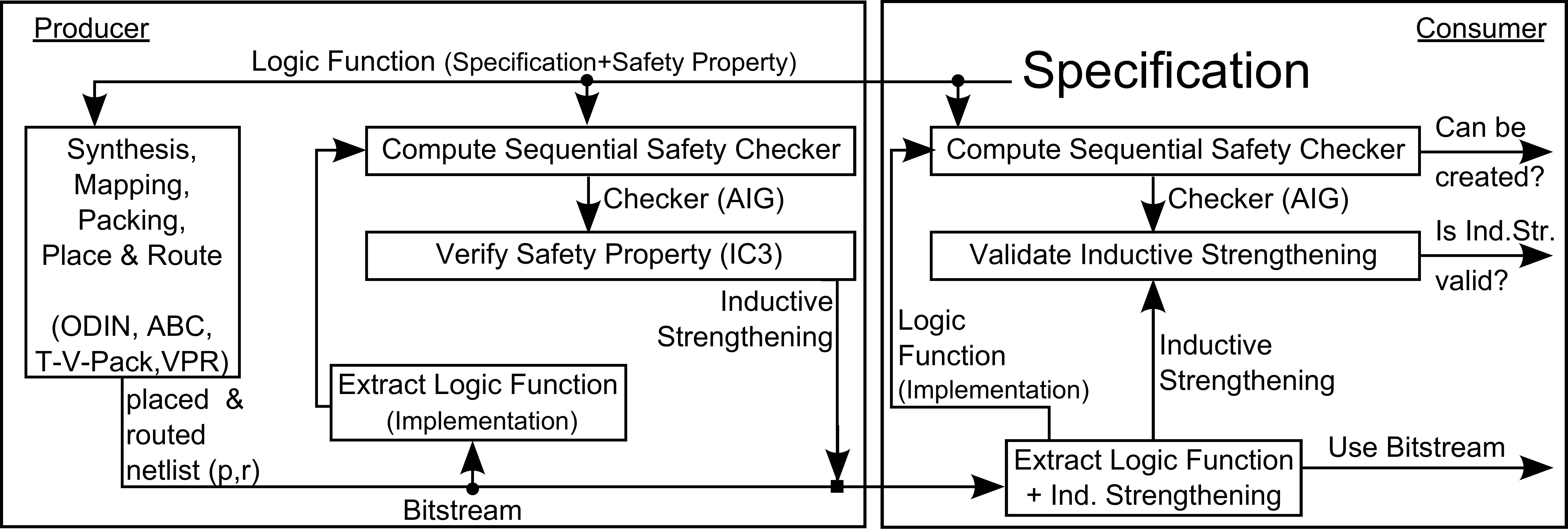}
		\caption{PCH workflow using IC3}
		\label{figure:Workflow}
\end{figure}

We start with explaining the overall workflow of our concept (see Figure \ref{figure:Workflow}), which structurally follows (but generalizes) the work in \cite{DP11}.

\paragraph{Overall workflow of PCH.} 
A consumer needs a particular hardware module for reconfiguration with some desired functionality.
Due to a lack of own resources, he employs a hardware producer for the synthesis.
The desired functionality is given as a logic specification  to the provider, and both parties agree upon a safety property that has to be met.
The provider thereafter produces the configuration in the form of a bitstream, e.g.\ by using a tool flow as described in \cite{DrzevitzkyBoundedSequEquiv}.

As the consumer lacks resources to verify safety himself, the producer furthermore generates a proof, which is easy to check for the consumer and convinces him of the safety. A PCH technique used in such a scenario should (a) cover the class of combinational and sequential circuits plus arbitrary safety properties, (b) should provide both provider and consumer with automatic procedures for proof generation and validation, respectively, and (c) should really allow for a proof validation significantly easier than proof generation ({\em speed-up}). 
For this part of the workflow, we propose the utilization of IC3 as proof generator and inductive strengthenings as proofs. This approach meets all three requirements. 

\begin{figure}
\begin{minipage}[hbt]{0.5\textwidth}
	\centering
	\includegraphics[width=0.9\textwidth]{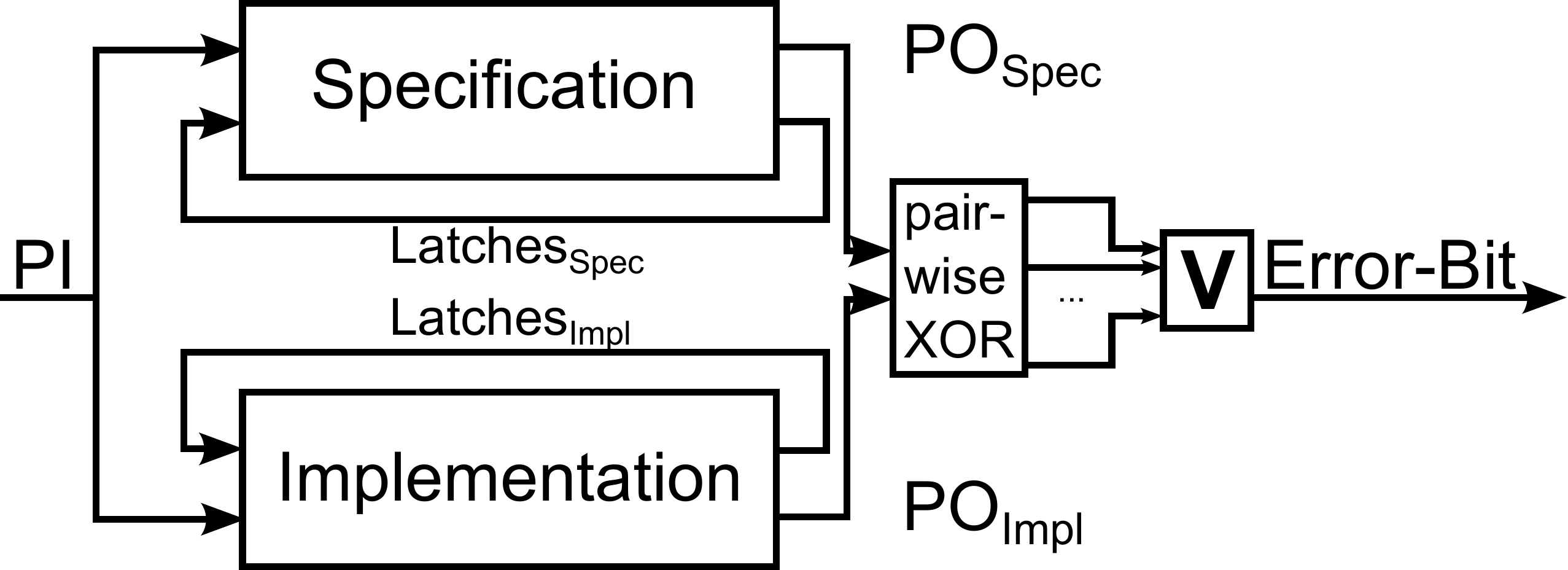}
		\caption{Sequential miter.}
		\label{figure:SequMiter}
\end{minipage}
\hfill
\begin{minipage}[hbt]{0.5\textwidth}
\centering
	\includegraphics[width=0.9\textwidth]{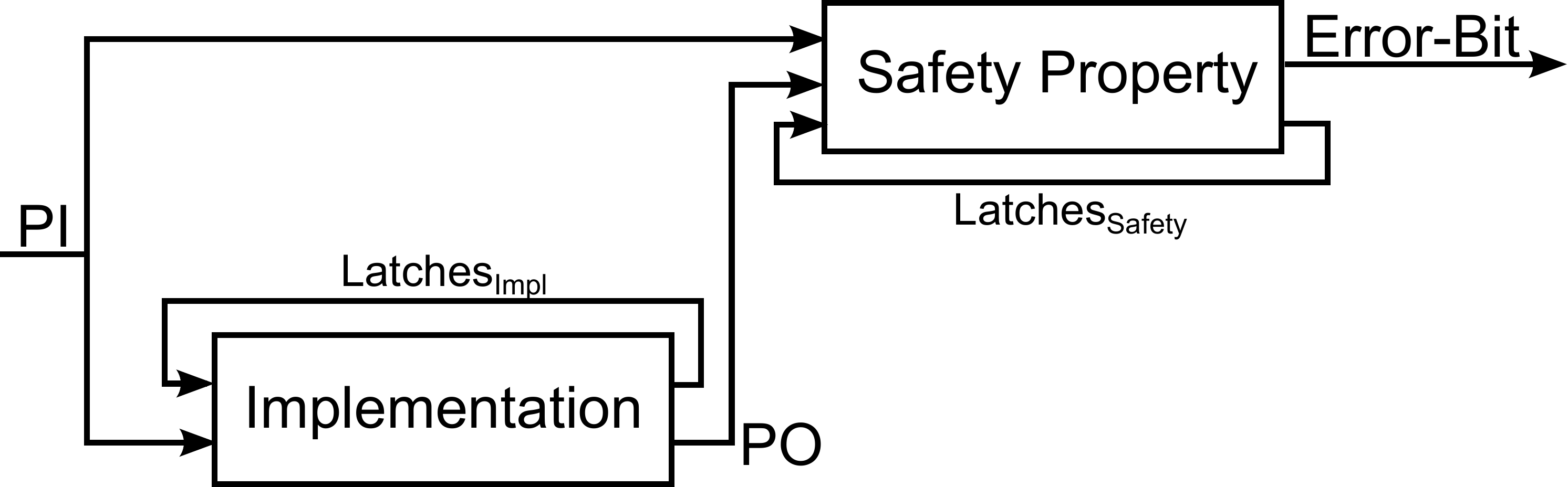}
		\caption{Generic sequential checker of a safety property.}
		\label{figure:GeneralSafetyProperty}
\end{minipage}
\end{figure}
As can be seen in Figure \ref{figure:Workflow}, in our approach the producer first combines the configuration to be shipped with the safety property into a sequential checker.
This can be a checker for full sequential equivalence, called {\em sequential miter} (Figure \ref{figure:SequMiter}), or in general any safety specification (Figure \ref{figure:GeneralSafetyProperty}).

Note, that these checkers are not plain combinational, but sequential, i.e., both the implementation and the safety property have their own latches for storing their states. 
In particular, no unrolling is necessary to create the checker as is the case in \cite{DrzevitzkyBoundedSequEquiv}.
This leads to significantly smaller circuits.

\paragraph{Proof generation and validation.}
The fact that we have sequential circuits here makes IC3 an ideal candidate for property checking. Moreover, we can readily use the inductive strengthening computed by IC3 as proof 
and ship it alongside the configuration.
The inductive strengthening is a boolean formula in CNF.
Given the boolean formulae $I$ and $T$, encoding the initial states and the transition relation of the sequential checker, and the safety property $P$ (most often encoded as an error bit of the checker), for each inductive strengthening $F$ the following queries are unsatisfiable:
\begin{align}
&\text{Initiation}: &I \wedge \neg F, \label{query:initiation}\\
&\text{Consecution}: &F \wedge T \wedge \neg F',\label{query:consecution}\\
&\text{Strengthening}: &F \wedge \neg P. \label{query:safety}
\end{align}

\noindent Note, that the primed formulae represent the next state variables (expressing the state after taking a transition).
Initiation and Consecution establish the inductiveness and the last property ensures the safety property.
For proof validation, the consumer builds his own checker to check the actually shipped configuration and proof against his own safety property.
This validation
is much faster than the generation on the producer's side and ensures that the shipped CNF formula is indeed an inductive strengthening.
Thus, the safety property holds and the consumer can safely reconfigure his hardware with the obtained configuration.

\paragraph{Tamper-proof PCH.}
We shortly explain why this concept is tamper proof.
As a premise, we require the miter generation and our tool {\it ISChecker} to be a trusted base. 
An attacker tampering with the shipped bitstream can either change the configuration, the proof or both.
If he changes the proof, the checker will show that it is not a valid inductive strengthening of the safety property regarding the shipped configuration.
If, by any chance, the changed CNF formula is still an inductive strengthening, this shows that there exists an over-approximation of the reachable states guaranteeing the property.
Thus, all reachable states adhere to the customer's safety property.
If the attacker changes the configuration, either no miter can be computed, as the number of inputs and outputs differ, or the inductive strengthening is not valid anymore.
If, by any chance, the inductive strengthening is still a valid one, the same argument as above holds.

\section{Implementation and Evaluation}

We evaluated our concept using several sequential miter created using AIGER\footnote{http://fmv.jku.at/aiger/} (safety properties other than functional equivalence require other tools).
A slightly changed IC3 reference implementation was utilized as generator of the PCH proofs, which were thereafter validated using our tool \textit{ISChecker}.

\paragraph{Implementation.} 
Whenever IC3 verifies that a safety property holds, it internally computes an inductive strengthening.
We slightly changed the reference implementation to store the found inductive strengthening as CNF in order to be shipped as a proof.
It is stored in a format similar to DIMACS, but using literals as in the AIGER-format, as is used in our validation tool.

Our tool {\it ISChecker} is based upon the IC3 reference implementation by using its integration of the tools AIGER and MiniSAT\footnote{http://minisat.se/}.
It reads the PCH proof $F$ as well as the AIG-file of the miter for which it constructs the CNF formulae $I$, $T$ and $P$. 
To check if the proof is a valid inductive strengthening of $P$, 
queries \ref{query:initiation},\ref{query:consecution} and \ref{query:safety} are issued to the SAT-solver.
Here, the negated CNF $\neg F$ can be handled in several ways, e.g., by using the Tseitin transformation or by splitting the query into several (since $\neg F$ is a disjunction of conjunctions).
Our evaluation of these approaches showed no significant difference in general, but for some small instances an overhead of the transformation.
If all queries are unsatisfiable, the PCH proof is indeed an inductive strengthening of $P$ for the checker.

\paragraph{Evaluation.}
For evaluation, we employed sequential multipliers of different size.
To this end, we created sequential miters using the tool AIGER to show functional equivalence of the implementation with a specification.
The results show a significant {\em speed-up} (Table \ref{table:runtimes}).
Thus, the main goals of PCH are met.

\begin{table}
	\begin{minipage}{0.5\textwidth}
\centering
		\scriptsize 
		\begin{tabular}{|l|c|c|}
					\hline
				Instance & IC3Gen & ISChecker \\
					\hline
					\hline
				sequ. 8Bit multiplier & 0.28 sec & 0.01 sec\\
					\hline
				sequ. 16Bit multiplier & 1.82 sec & 0.01 sec\\
					\hline
				sequ. 32Bit multiplier & 55.18 sec & 0.04 sec\\
					\hline
				sequ. 64Bit multiplier & 664.6 sec & 0.11 sec\\
					\hline
				sequ. 128Bit multiplier & 73910 sec & 0.17 sec\\
					\hline
		\end{tabular}
	\end{minipage}
	\begin{minipage}{0.5\textwidth}
	\centering
		\scriptsize
		\begin{tabular}{|l|r|r|}
					\hline
				Instance & IC3Gen & ISChecker\\
					\hline
					\hline
				6s8 & 395.23 sec & 0.82 sec\\
					\hline
				6s206rb025 &  2.92 sec & 1.11 sec\\
					\hline
				6s310r &  249.95 sec & 0.37 sec\\
					\hline
				6s313r & 29.00 sec & 7.06 sec\\
					\hline
				6s364rb03158 & 512.93 sec & 22.52 sec\\
					\hline
		\end{tabular}
	\end{minipage}
		\caption{Proof generation (IC3Gen) vs.\ validation (ISChecker) using sequential multiplier benchmarks (left) and HWMCC 2013 benchmarks (right).}
		\label{table:runtimes}
\end{table}

Furthermore, our approach has several advantages over previous approaches, e.g.,
the sizes of our proofs
are significantly smaller (less than 1.5KB) than in the resolution based unrolling approach \cite{DrzevitzkyBoundedSequEquiv} (up to 700MB)
and the creation of the miter itself is faster, because no unrolling needs to be computed.
Further evaluation using several single track benchmarks from the hardware model checking competition 2013 emphasizes the significant {\em speed-up} in validation (Table \ref{table:runtimes}).

Our approach rules out the need to consider more than one transition step and is therefore only reasonably applied to sequential circuits.
Note, that with increasing complexity of the transition relation our validation slows down due to query \ref{query:consecution}.
In these cases, we propose to speed up our validation process by using an additional proof for said query in the form of an unsatisfiability-trace.
\section{Conclusion}

In this paper, we extended the existing PCH approach of Drzevitzky \cite{DrzevitzkyBoundedSequEquiv} to handle fully sequential circuits.
We proposed IC3 as a proof generator for these circuits, and showed how proof validation works. 
We presented an experimental evaluation that proof validation is significantly faster than proof generation. Furthermore, our approach remains tamper-proof, which is crucial for PCH. 

\paragraph{Related Work.} Proof-carrying hardware was proposed by Drzevitzky et al. \cite{DP11,DrzevitzkyBoundedSequEquiv}.
Her approach utilizes the trace of a resolution-based unsatisfiability proof as PCH proof. The technique handles combinational circuits and bounded unfoldings of sequential circuits only, and checks full functional equivalence with a specification only.  
Our approach uses a similar work flow, but heavily differs in being able to check arbitrary safety properties of sequential circuits. 
In addition, the employed proofs as well as their generators and checkers differ completely.

Other work regarding proof-carrying hardware also handles more general properties than simply functional equivalence.
Love et al.\ \cite{LoveIP} introduced a framework in which proofs in the Coq proof assistant language can be generated from a subset of Verilog. However, no evaluation of their approach is given, and furthermore, the approach is not fully automatic as it uses an interactive prover. 

\medskip
\noindent {\bf Acknowledgement.} We thank the research group of Marco Platzner for providing us with the multiplier configurations.  

\bibliographystyle{plain}
\bibliography{references}

\begin{thebibliography}{1}

\bibitem{Bradley11}
Aaron~R. Bradley.
\newblock Sat-based model checking without unrolling.
\newblock In Ranjit Jhala and David~A. Schmidt, editors, {\em VMCAI}, volume
  6538 of {\em Lecture Notes in Computer Science}, pages 70--87. Springer,
  2011.

\bibitem{DP11}
S.~Drzevitzky and M.~Platzner.
\newblock Achieving hardware security for reconfigurable systems on chip by a
  proof-carrying code approach.
\newblock In {\em Reconfigurable Communication-centric Systems-on-Chip
  (ReCoSoC), 2011 6th International Workshop on}, pages 1--8, 2011.

\bibitem{DrzevitzkyBoundedSequEquiv}
Stephanie Drzevitzky.
\newblock {\em Proof-carrying hardware : A novel approach to reconfigurable
  hardware security}.
\newblock Dissertation, University of Paderborn, 2012.

\bibitem{LoveIP}
Eric Love, Yier Jin, and Yiorgos Makris.
\newblock Proof-carrying hardware intellectual property: A pathway to trusted
  module acquisition.
\newblock {\em IEEE Transactions on Information Forensics and Security},
  7(1):25--40, 2012.

\bibitem{Necula97}
George~C. Necula.
\newblock Proof-carrying code.
\newblock In Peter Lee, Fritz Henglein, and Neil~D. Jones, editors, {\em POPL},
  pages 106--119. ACM Press, 1997.

\end{thebibliography}

\end{document}